\documentclass[sigconf]{acmart}
\setkeys{acmart.cls}{balance=false}

\AtBeginDocument{%

}
\AtBeginDocument{\sloppy}

\usepackage{algorithm}
\usepackage{algpseudocode}
\usepackage{threeparttable}
\usepackage{tabularx}
\usepackage{array}
\usepackage[table]{xcolor}
\newcolumntype{Y}{>{\raggedright\arraybackslash}X}

\acmConference[CAIS~'26]{ACM Conference on AI and Agentic Systems}{May 26--29, 2026}{San Jose, CA, USA}
\acmYear{2026}
\copyrightyear{2026}

\title[CAMI:Cost-Aware Multi-Indexing]{CAMI: Practical Cost-Aware Agent-Guided Multi-Indexing for Semantic Retrieval}
\begin{document}

\author{Adnan Qidwai}
\email{Adnan.Qidwai1@ibm.com}
\orcid{0009-0009-2181-287X}
\affiliation{%
  \institution{IBM Software Innovation Lab}
  \country{India}
}

\author{Anand Eswaran}
\email{Anand.Eswaran@ibm.com}
\orcid{0009-0004-2189-555X}
\affiliation{%
  \institution{IBM Software Innovation Lab}
  \country{India}
}

\author{Sonam Mishra}
\email{Sonam.Gupta7@ibm.com}
\orcid{0000-0003-0562-8887}
\affiliation{%
  \institution{IBM Software Innovation Lab}
  \country{India}
}

\author{Jaydeep Sen}
\email{jaydesen@in.ibm.com}
\orcid{0009-0009-9202-0300}
\affiliation{%
  \institution{IBM Software Innovation Lab}
  \country{India}
}

\author{Sachindra Joshi}
\email{jsachind@in.ibm.com}
\orcid{0009-0009-8089-9493}
\affiliation{%
  \institution{IBM Software Innovation Lab}
  \country{India}
}

\renewcommand{\shortauthors}{A. Qidwai et al.}

\begin{abstract}
 RAG ingestion pipelines frequently augment search corpus index with semantic enrichment indices (e.g., synthetic queries or summaries generated from corpus chunks) that are subsequently queried alongside the base index to improve retrieval via better alignment between document representations and user intent. While these supplementary representations substantially improve retrieval quality, they introduce a computational bottleneck: the configuration space of enrichment types and generator models is combinatorial, and the cost of exhaustive index-time evaluation scales linearly with corpus size. We introduce CAMI (Cost-Aware Multi-Indexing), a framework that formalizes multi-index construction as a budgeted, multi-objective portfolio selection problem. CAMI targets the upstream decision of which enrichment views to generate and materialize before the retrieval backend is applied. CAMI incorporates three primary mechanisms: (i) an agentic discovery phase that proposes corpus-specific representation templates; (ii) an atomic-unit search procedure that evaluates individual enrichment-model pairs and recombines them via fidelity-local closure to identify synergistic portfolios; and (iii) a confidence-aware promotion schedule that prunes unpromising configurations early, decoupling optimization spend from total corpus size. We evaluate CAMI across diverse retrieval corpora. Our findings reveal that the framework systematically isolates high-recall portfolios under strict budget constraints, outperforming standard content-only baselines in challenging settings by up to \textcolor{black}{9.4\% recall@10}. Further, CAMI is able to systematically identify these high-recall portfolios using up to \textcolor{black}{5x} less budget compared to random search baselines, making our approach practical in real production scenarios.
\end{abstract}




\begin{teaserfigure}
    \centering
    \includegraphics[width=\linewidth]{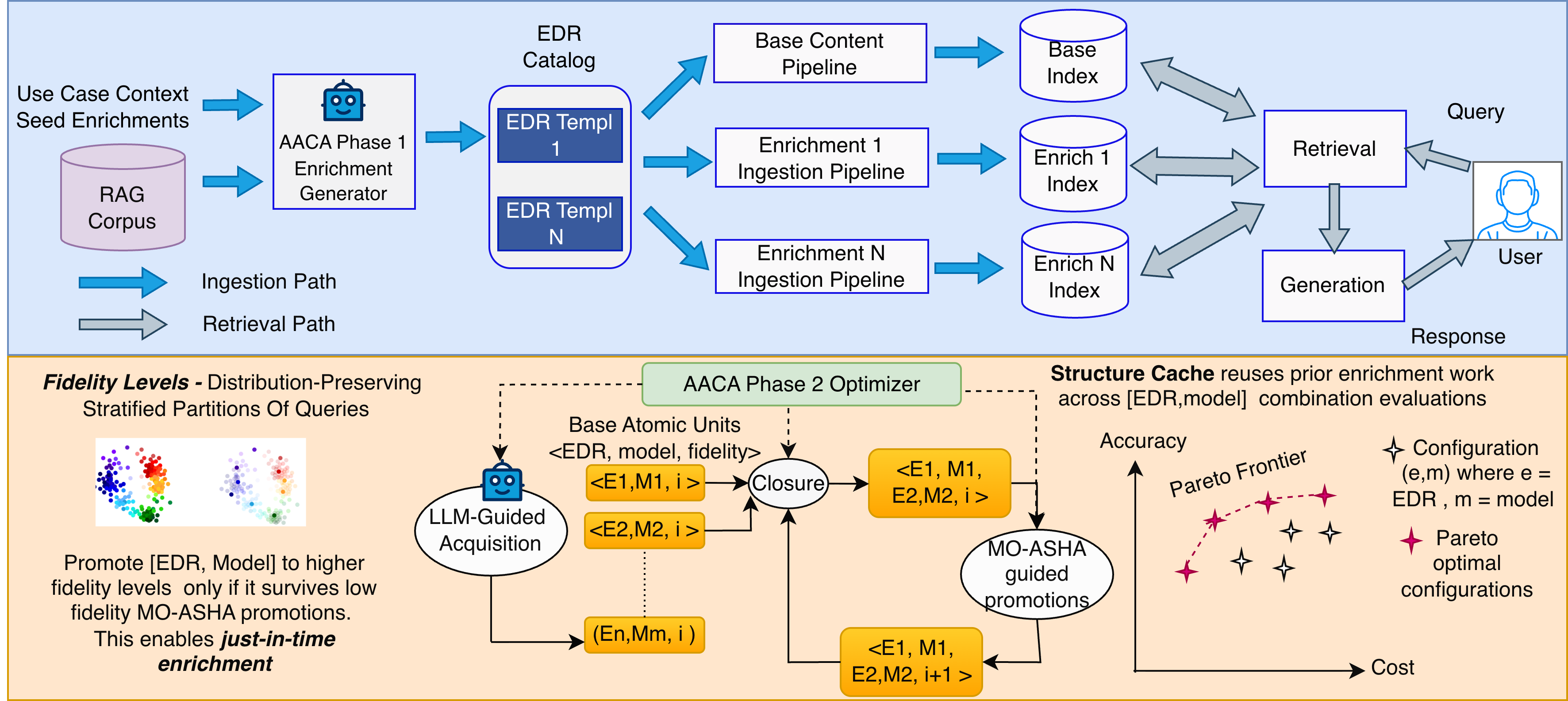}
    \caption{\textbf{The CAMI Framework.} The top panel outlines the multi-index ingestion and retrieval pipeline, where an agent proposes dataset-specific Enriched Data Representations (EDRs) to be queried alongside the base text. The bottom panel details the cost-aware search engine. To avoid expensive full-corpus evaluations, CAMI tests atomic units (EDR--model pairs) on small, representative query subsets (fidelities). Successful units are recombined into multi-index portfolios (closure) and promoted to larger evaluation subsets using MO-ASHA. By caching prior work, the system efficiently identifies optimal configurations on the recall--cost Pareto frontier.}
    \label{fig:method-overview}
\end{teaserfigure}

\acmYear{2026}\copyrightyear{2026}
\setcopyright{cc}
\setcctype[4.0]{by}
\acmConference[ACM CAIS '26]{ACM Conference on AI and Agentic Systems}{May 26--29, 2026}{San Jose, CA, USA}
\acmBooktitle{ACM Conference on AI and Agentic Systems (ACM CAIS '26), May 26--29, 2026, San Jose, CA, USA}
\acmDOI{10.1145/3786335.3813171}
\acmISBN{979-8-4007-2415-2/26/05}

\maketitle

\section{Introduction}
\label{sec:intro}

\begin{figure*}[t]
    \centering
    \includegraphics[width=0.95\textwidth]{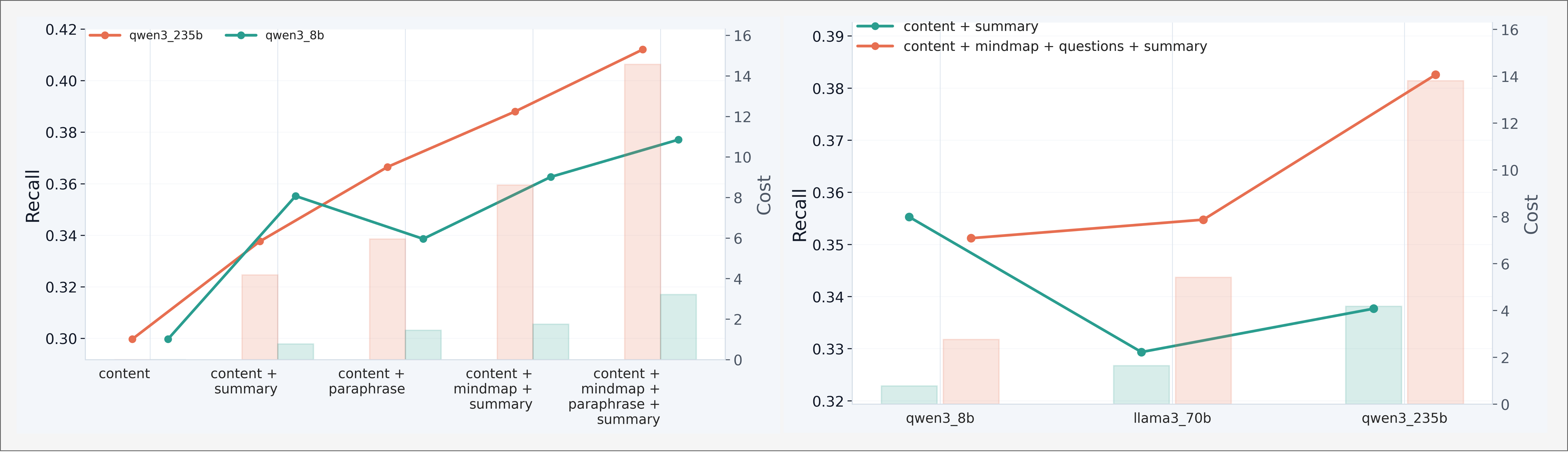}
    \caption{\textbf{The Recall--Cost Trade-off.} This figure illustrates how different multi-index combinations impact both retrieval quality (recall, shown as lines on the left y-axis) and structural cost (shown as bars on the right y-axis) on the \textit{StatCanDialogueRetrieval} dataset. \textbf{Left:} Adding more representations to the index while keeping the generator model fixed generally increases recall, but stacks construction costs. \textbf{Right:} Upgrading to different generator models (e.g., moving from \textit{Qwen3-8B} to \textit{Qwen3-235B}) for a fixed configuration may not always yield better retrieval accuracy but it drastically inflates the token spend. This demonstrates the necessity of cost-aware optimization.}
    \label{fig:cost_recall_tradeoff}
\end{figure*}

\paragraph{The Promise of Multi-Index Retrieval.} Modern retrieval stacks increasingly rely on index-time enrichment to mitigate a recurrent failure mode in retrieval-augmented generation: lexical, stylistic, and structural mismatch between user queries and source text. A single representation is often brittle when the corpus expresses concepts differently from the query's phrasing. To address this, systems like EnrichIndex  \cite{chen2025enrichindexusingllmsenrich} and MetaData-As-Data\cite{yousuf2026utilizingmetadatabetterretrievalaugmented} propose static Enriched Data Representations (EDRs), which are auxiliary textual view generated via large language models (LLMs) from raw chunks of a corpus and index these as supplementary indices during the data ingestion phase. For example, an EDR might be a selective paraphrase highlighting some theme of interest, a concise objective summary, or a list of synthetic questions generated against each chunk in a RAG corpus. At query time, the system retrieves across multiple EDR indices and fuses their rankings across indices using rankers like RRF (reciprocal rank fusion) \cite{10.1145/1571941.1572114}. This multi-index approach improves recall of semantic retrieval by increasing the probability that at least one EDR aligns with the user's query intent, while preserving a simple query-time control flow that avoids expensive per-query language model inference approaches such as agentic RAG  \cite{singh2025agenticretrievalaugmentedgenerationsurvey}. In  Figure \ref{fig:cost_recall_tradeoff} (left), we show the \textcolor{black}{recall@10} gains associated with individual EDRs enrichments (content+summary, content+paraphrases) and portfolios of EDRs (content + mindmap + summary, content+mindmap + paraphrase + summary) for the StatCanDialogueRetrieval \cite{Lu_2023} data set. We also show in Figure \ref{fig:cost_recall_tradeoff} (right) that the impact of model selection on recall/accuracy is non-monotonic with performance varying across EDRs (content + summary, content + mindmap + questions + summary) rather than scaling commensurately with model cost. The same figure also motivates why CAMI searches the joint EDR-model space: a low-cognitive-load EDR may be well served by a cheap model, while a more demanding representation can have dataset-dependent model requirements that cannot be assumed away in advance.

\paragraph{The Combinatorial Cost Bottleneck.} As Figure \ref{fig:cost_recall_tradeoff} demonstrates, the same mechanism that improves recall also creates a strict systems bottleneck. Every additional EDR type and generator model choice incurs offline token spend, storage overhead, and evaluation cost. Before one even decides between retrieval backends or physical index layouts, the system must decide which views are worth generating at all. The number of candidate portfolios grows combinatorially, rendering naive enumeration financially and computationally unaffordable. The design objective is therefore inherently multi-objective: practitioners require a frontier of non-dominated recall--cost choices under strict optimization budgets.

\paragraph{The Limits of General-Purpose Sampling.} Although black-box multi-objective optimizers such as Multi-Objective Bayesian Optimization can explore this space, they treat configurations as monolithic, path-independent points. This abstraction mismatches multi-index retrieval, where portfolios are composed from reusable atomic units. For instance, two portfolios sharing a summary generated by \textit{Qwen3-8B} \cite{yang2025qwen3} are evaluated as independent trials despite overlapping construction and retrieval evidence. In practice, once an EDR–model pair is materialized, its outputs and cached results can be reused across many portfolios at negligible marginal cost. Ignoring this structure induces redundant evaluation spend and slows frontier discovery, motivating an explicitly atom-aware optimization strategy that supports both atomic discovery and systematic recombination.

\paragraph{Agentic EDR Discovery.} Existing pipelines assume a fixed set of enrichment templates, yet the representations that close recall gaps are corpus- and task-dependent. Static catalogs therefore risk accumulating construction cost without addressing dominant mismatch modes. We observe that instead enrichment design can be treated as a use-case data-set conditioned budgeted search problem: an agent proposes corpus-conditioned EDR templates from seed query-evidence pairs and corpus context, which are validated and retained only when they improve the recall–cost frontier. This reframes EDR creation from a one-time heuristic choice into an evidence-driven, corpus-adaptive step aligned with the same optimization objective as portfolio selection.

\paragraph {The CAMI Framework.} We introduce CAMI, an atomic search and discovery framework over the state space of multi-index EDR-model combinations, thus efficiently identify combinations on the Pareto frontier within a budget constraint. \textit{The combinations identified by CAMI are then used for enrichment at scale over the larger corpus.} CAMI takes the shape of a two-phase, budget-constrained optimization framework that solves multi-index selection via multi-fidelity, atom-aware search. A first phase performs agentic EDR discovery, generating corpus-specific EDR prompt templates that are validated and frozen under budget accounting so that the downstream search space remains stationary and auditable. The next phase implements AACA - Atomic-unit Acquisition + Closure + MO-ASHA (Multi-Objective Asynchronous Successive Halving) \cite{schmucker2021multiobjectiveasynchronoussuccessivehalving} over atomic units (EDR-model pairs) : it bootstraps with cheap baselines, incrementally acquires atomic units using an LLM-based agentic process, and exploits their compositional structure through fidelity-local closure, recombining validated atoms to form new portfolios instead of sampling monolithic configurations. Candidate portfolios are evaluated on nested query subsets (fidelities) to bound cost, while MO-ASHA promotions with UCB(Upper-Confidence-Bound)-style ranking and frontier gating allocate higher-fidelity budget only to configurations likely to improve the Pareto frontier. Additional mechanisms—drift-aware challenger lanes, starvation guards, and retrieval/result caching—stabilize exploration and prevent wasted spend. Overall, CAMI operationalizes a practical cost-aware, compositional search that concentrates evidence on promising portfolios while rigorously tracking budget and recall–cost trade-offs. Figure 1 provides an overview of our approach.

\paragraph{Core Research Questions.} To evaluate the efficacy of this decomposition, we address the following foundational questions: \textbf{Are existing multi-index retrieval pipelines fundamentally limited by offline construction costs, and can an atomic-unit search procedure reliably isolate Pareto-optimal configurations under strict budget constraints?} Furthermore, \textbf{Does agent-guided template discovery provide measurable recall improvements over static enrichment strategies without bypassing rigorous cost accounting?} A solution to these research questions can fundamentally change how multi-index retrieval systems are designed and deployed today, potentially producing much more efficient and effective strategies tied to specific use-cases. 

\noindent\textbf{Summary of Contributions.} This paper makes the following contributions:
\begin{enumerate}
    \item We formulate retrieval enrichment as a budgeted Pareto-frontier discovery problem with explicit cost semantics, separating construction cost from the cost of choosing and executing an EDR combination.
    \item We introduce the AACA optimizer, which bridges MO-ASHA budget controls with fidelity-local closure to exploit the computational reuse of EDR-model pairs during search.
    \item We integrate a phase of agentic EDR discovery via LLMs into the same workflow, enabling purpose-crafted enrichment strategies tailored to the dataset and use case, while obeying the budget constraints.
    \item We operationalize CAMI as a practical cost-aware multi-index RAG ingestion pipeline using our techniques at its core and evaluate it comprehensively on challenging public datasets such as StatCanDialogueRetrieval\cite{Lu_2023}, Bright \cite{su2025brightrealisticchallengingbenchmark} and NQTables \cite{herzig2021opendomainquestionanswering}. CAMI consistently identifies cost-optimal EDR–model portfolios that improve recall by up to \textcolor{black}{9.4\%} with search budget savings of \textcolor{black}{5x} relative to random search baselines.
\end{enumerate}

\section{Related Work}
\label{sec:related}

\paragraph{Retrieval Index Enrichment and Multi-Indexing.} Modern RAG systems leverage richer document representations to improve recall. For example, EnrichIndex \cite{chen2025enrichindexusingllmsenrich} uses an LLM offline to append semantic annotations (e.g. summaries, inferred queries) to each document, boosting recall significantly. MC-Indexing \cite{dong2024multiviewcontentawareindexinglong} propose keyword and summary based views for long documents. Other approaches use multiple chunk views or embeddings. HeteRAG \cite{yang2025heteragheterogeneousretrievalaugmentedgeneration} decouples retrieval and generation indexes by storing short chunks for generation and longer, context-rich chunks for retrieval, yielding substantial gains in both effectiveness and efficiency. Similarly, graph-based RAG methods \cite{peng2024graphretrievalaugmentedgenerationsurvey,edge2025localglobalgraphrag, gutierrez2025hipporagneurobiologicallyinspiredlongterm, gutierrez2025ragmemorynonparametriccontinual, Ghassel_2025, lazygraphrag} employ graph-structured indexes, while RAPTOR  \cite{sarthi2024raptorrecursiveabstractiveprocessing} and SIRE-RAG \cite{zhang2025sireragindexingsimilarrelated} build hierarchical summary trees to compress detailed context trading reasoning for efficiency. The KET-RAG \cite{huang2025ketragcostefficientmultigranularindexing} framework explicitly constructs a small “skeleton” graph over key chunks and a lightweight bipartite index for all chunks, achieving comparable QA performance to full Graph-RAG with orders-of-magnitude lower indexing cost. These works illustrate that using multiple, multi-granular representations (e.g. vectors or graphs) can improve retrieval quality while managing cost. Multi-vector retrieval methods such as \cite{Scheerer_2025, dhulipala2024muveramultivectorretrievalfixed,santhanam2022plaidefficientenginelate} build on the early promise of Colbert \cite{khattab2020colbertefficienteffectivepassage,santhanam2022colbertv2effectiveefficientretrieval} making such techniques increasingly practical: they use optimized pipelines to cut end-to-end multi-vector query latency while preserving accuracy. In summary, prior work demonstrates the power of multi-index or multi-vector retrieval. However, existing methods employ fixed augmentation schemes or indexes; none automatically search the space of possible semantic enrichments under a cost budget. CAMI instead dynamically discovers and selects among many enrichment templates to optimize the recall–cost tradeoff.

\paragraph{Cost-Aware Optimization in Compound AI Pipelines.} The compound AI community has increasingly explored joint optimization of accuracy and cost for data pipelines including enrichments. For semantic data pipelines, Abacus \cite{abacus} introduces a cost-based optimizer for “semantic operators” (LLM-powered maps, filters, etc.) by using a novel Pareto-Cascades search. MOAR (Multi-Objective Agentic Rewrites) \cite{wei2026multiobjectiveagenticrewritesunstructured} extends this to DocETL \cite{shankar2025docetlagenticqueryrewriting} pipelines by adding dozens of rewrite directives (e.g. decomposing filters) and using a global search with a multi-armed bandit to optimize both cost and accuracy. In the RAG setting, Barker et al. \cite{barker2025fastercheaperbettermultiobjective} apply multi-objective Bayesian optimization over LLM, embedding, and retriever choices, finding Pareto-optimal configurations of cost and quality. Syftr \cite{conway2025syftrparetooptimalgenerativeai} searches over broad RAG pipeline designs (including agents like verifiers and rerankers) with multi-objective Bayesian methods and early-stopping: it locates Pareto-frontier flows that are ~9× cheaper with minimal accuracy loss. Systematic hyperparameter optimization for RAG parameter configurations is explored in \cite{orbach2025analysishyperparameteroptimizationmethods}.  These systems underscore that explicit cost–quality tradeoffs can be managed via Pareto-aware search. CAMI builds on this by treating multi-index construction as a multi-objective selection problem, tracing the recall–cost frontier under strict budgets. In contrast to these approaches, we leverage the sub-structure of the state space to reduce budget-intensive combinatorial explorations.

\paragraph{LLM-Guided Search and Model Selection in Compound Systems.} Prior work has shown that leveraging LLMs themselves as exploration agents \cite{llmbeamsearch,liu2025decisiontreeinductionllms,ye2024reevolargelanguagemodels} or evaluators \cite{gu2025surveyllmasajudge, mahdavi2025scalinggenerativeverifiersnatural, yang2025critiqueverifyaccuratehonest, kim2024prometheusinducingfinegrainedevaluation, chiang2024chatbotarenaopenplatform, zeng2024evaluatinglargelanguagemodels} can efficiently navigate complex system configurations. GEPA \cite{agrawal2026gepareflectivepromptevolution} uses LLM reflection to iteratively propose and mutate prompts, combining Pareto-optimal variations; this technique can be adapted to propose candidate combinations. For planning tasks, CATS \cite{zhang2026costawarenesstreesearchllmplanning} introduces a cost-aware Monte Carlo Tree Search using LLM-assisted planning to prune unpromising parts of search space under strict budgets. These methods illustrate powerful patterns: LLMs can autonomously explore design spaces (prompts, models, actions) with informed search. CAMI's AACA phase adopts a similar philosophy by using an LLM “agent” loop to propose corpus-specific portfolio candidates from candidate atoms based on historical information coupled with knowledge of data set / use case details to explore the combinatorial index configuration space under cost constraints. Model selection for single task settings is covered in \cite{chen2023frugalgptuselargelanguage, ramirez2024optimisingcallslargelanguage, shekhar2024optimizingcostsllmusage, stripelis2024tensoroperaroutermultimodelrouter}. LLMSELECTOR \cite{chen2025optimizingmodelselectioncompound} studies model selection in multi-module systems. In contrast, CAMI co-optimizes EDR selection jointly with model selection and reuses per-model work efficiently across portfolios during search exploration.

\section{Problem Formulation \& Cost Semantics}
\label{sec:problem}

\begin{figure}[t]
    \centering
    \includegraphics[width=\columnwidth]{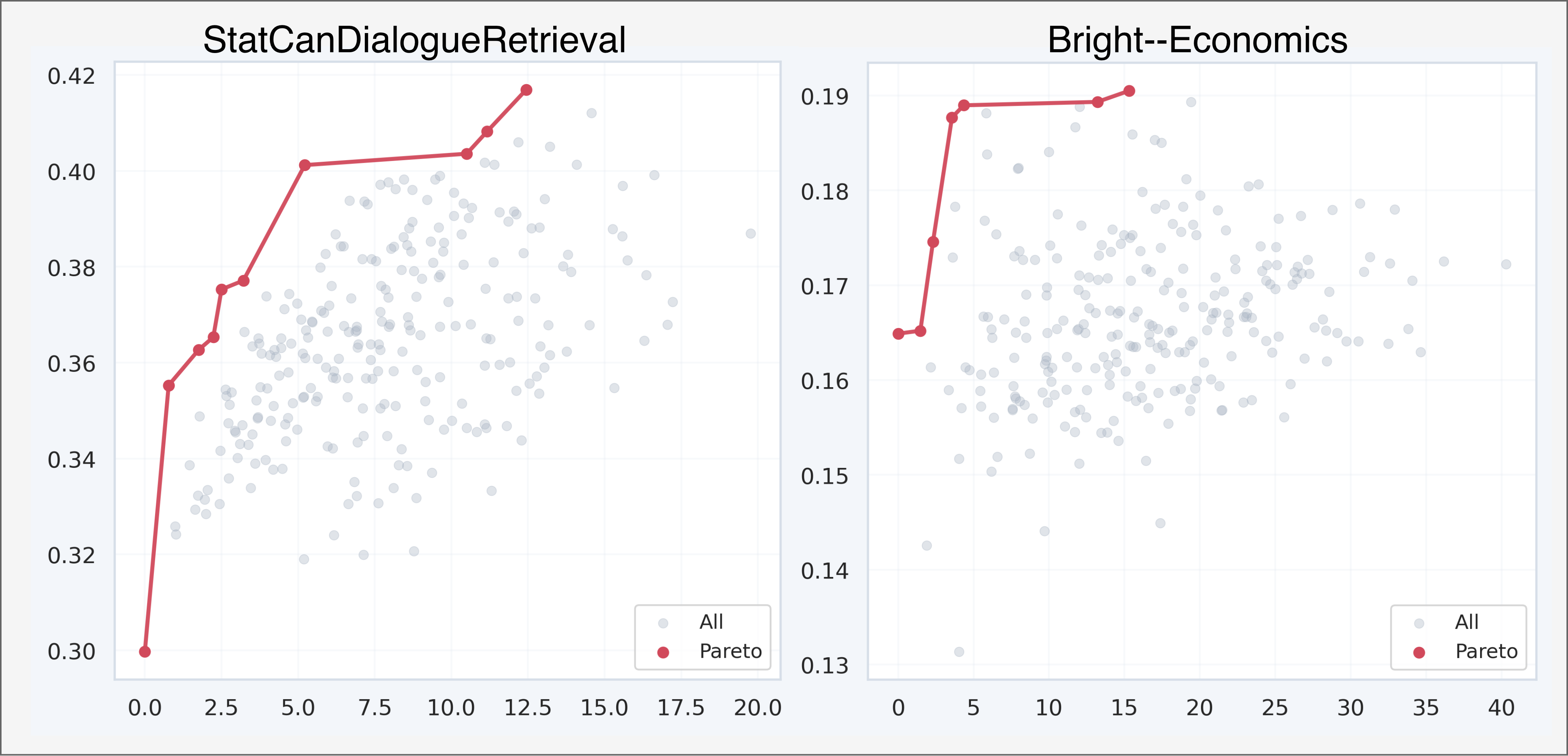}
    \caption{\textbf{The Empirical Pareto Frontier.} Scatter plots illustrating the trade-off between structural cost (x-axis) and retrieval recall (y-axis) across evaluated multi-index combinations. The gray dots represent individual combinations tested during the search phase. The red line traces the empirical Pareto frontier, isolating the optimal combinations that maximize recall for any given deployment budget. The dense cloud of suboptimal gray combinations visually emphasizes the necessity of an efficient search algorithm.}
    \label{fig:pareto_frontier}
\end{figure}

\paragraph{CAMI Objective}CAMI's objective is to explore the state space of multi-index EDR-model combinations and efficiently identify those portfolio combinations on the Pareto frontier. These identified combinations are then used as supplementary indices over the larger RAG corpus coupled with multi-index retrieval from these supplementary indices during query-time. The state space of portfolio combinations \textcolor{black}{spanning enrichments \textit{[summaries,paraphrases,questions,keywords]} and models [\textit{qwen3\_8b}, \textit{qwen3\_235b}, \textit{llama3\_70b}] } \cite{yang2025qwen3, grattafiori2024llama} evaluated using Grid Search are shown in Figure \ref{fig:pareto_frontier} for StatCanDialogueRetrieval and Bright-Economics datasets with the Pareto frontier portfolio combinations marked in red. To optimize multi-index construction, we formalize the search space of enrichment combinations, the optimization objective, and the dual nature of computational costs. Table~\ref{tab:notation} summarizes the core notation used throughout the framework.

\paragraph{The Multi-Index Combination Space.} We consider a retrieval corpus of $N$ chunks and a benchmark query set $Q$ containing ground-truth relevance labels. Let $\mathcal{E}$ denote the catalog of available enrichment types, and $\mathcal{M}_e$ the set of candidate generator models for enrichment $e \in \mathcal{E}$. An atomic unit $a = (e, m)$ is defined as a specific enrichment type paired with a specific generator model. A multi-index combination is strictly defined as a set of active atomic units. \textit{For example, a combination might consist of (\textit{Summary}, \textit{Llama-3.3-70B}) and (\textit{Synthetic Questions}, \textit{Qwen3-8B}), which are implicitly fused with the base content index at query time.}

\paragraph{Explicit Cost Semantics.} \label{cost-sem} To properly solve this bottleneck, the system must separate the theoretical cost of deploying a final index from the actual dollar spend incurred during the search process. This decoupling is a core idea of our framework, conceptually illustrated in Figure~\ref{fig:method-overview}. We track two complementary cost ledgers:
\begin{enumerate}
    \item \textbf{Runtime Uncached Spend ($C_{\mathrm{uncached}}$):} This ledger tracks newly incurred, real-world dollar costs during the search phase. It includes cache-miss enrichment generation and language model policy calls. This metric strictly governs the search optimization budget limit $B_{\mathrm{search}}$.
    \item \textbf{Structural Cost ($C_{\mathrm{struct}}$):} This is a cache-agnostic accounting mechanism used to evaluate the final deployment cost of a combination. It charges each atomic unit within a combination by its nominal construction cost. This prevents double-counting cached units and ensures fair comparisons across combinations, accurately representing the true cost to deploy the final index from scratch.
\end{enumerate}
See Appendix~\ref{app:budget-token-cost} for the exact token-pricing model. \textit{We also note that using prefix-caching across various EDR generations can also save cost spent towards constructing the atomic units. We could not evaluate saving metrics over KV-Caching because of lack of access to inference infrastructue.}

\paragraph{The Evaluation Cost Bottleneck.} Discovering the optimal multi-index combination is financially prohibitive because generating textual outputs over $N$ chunks for every proposed atomic unit incurs massive LLM token costs. Let $\rho\in(0,1]$ denote the fraction of chunks evaluated during search, $K$ the total number of distinct atomic units tested, and $n_{\mathrm{sel}}$ the number of units kept in the final deployed combination. The search space exploration cost (SSEC) scales as $\Theta(K \rho N)$, while the one-time deployment cost (EC) scales as $\Theta(N n_{\mathrm{sel}})$. Consequently, when $\rho K / n_{\mathrm{sel}} > 1$, searching for the best combination costs more than simply building it. This necessitates separating the theoretical cost of deploying a final index from the actual dollar spend incurred during search.

% \paragraph{Search vs. Construction Cost.} The fundamental problem is that search exploration frequently costs more dollars than final construction. Let $\rho\in(0,1]$ denote the fraction of chunks evaluated during search, let $K$ be the total number of distinct atomic units tested, and let $n_{\mathrm{sel}}$ denote the number of atomic units kept in the final deployed combination. The search space exploration cost (SSEC) scales with the number of chunk-level constructions performed during the search phase:
% \begin{equation}
% \mathrm{SSEC}=\Theta(K \rho N).
% \end{equation}
% By contrast, the one-time deployment enrichment cost (EC) for the final selected combination scales as:
% \begin{equation}
% \mathrm{EC}=\Theta(N n_{\mathrm{sel}}).
% \end{equation}
% When $\rho K / n_{\mathrm{sel}} > 1$, searching for the best combination requires more computational spend than simply building it once it is known. 

\paragraph{Formal Optimization Problem.} The formal optimization problem is to discover a Pareto frontier of multi-index combinations under a strict cost budget. The search target is the Pareto set of combinations that maximize empirical retrieval recall ($R$) and minimize structural deployment cost ($C_{\mathrm{struct}}$). To handle empirical noise bounded by finite query samples, we define dominance using small slack tolerances $\varepsilon_R \ge 0$ and $\varepsilon_C \ge 0$. A combination $c'$ dominates $c$ (denoted $c' \succ c$) if it performs better or equal across both dimensions up to the tolerance, with at least one strictly superior metric:
\begin{equation}
\begin{aligned}
c' \succ c \iff\ &
R(c') \ge R(c)-\varepsilon_R\ \wedge\ C_{\mathrm{struct}}(c') \le C_{\mathrm{struct}}(c)+\varepsilon_C \\
& \wedge\ \big(R(c') > R(c)+\varepsilon_R\ \vee\ C_{\mathrm{struct}}(c') < C_{\mathrm{struct}}(c)-\varepsilon_C\big).
\end{aligned}
\end{equation}
The ideal Pareto set is $\mathcal{P}=\{c:\nexists c' \text{ s.t. } c' \succ c\}$. Because full-dataset enumeration is impossible, our objective is to find an empirical approximation $\hat{\mathcal{P}}$ subject to the hard constraint that the sum of runtime uncached dollar spend across all search steps $t$ does not exceed the allowed optimization budget:
\begin{equation}
\sum_{t=1}^{T} C_{\mathrm{uncached}}(c_t) \le B_{\mathrm{search}}.
\end{equation}

\begin{table}[t]
\caption{Core notation.}
\label{tab:notation}
\centering
\scriptsize
\setlength{\tabcolsep}{4pt}
\renewcommand{\arraystretch}{1.05}
\begin{tabular}{@{}lp{0.82\columnwidth}@{}}
\toprule
Symbol & Meaning \\
\midrule
$N$ & Number of corpus chunks (the indexed units). \\
$Q$ & Full benchmark/query-log set with gold relevant document IDs per query. \\
$\ell\in\{0,1,\dots,L\}$ & Fidelity rung (nested query subsets). \\
$Q_\ell$ & Query subset at fidelity $\ell$ with $Q_0\subset Q_1\subset \dots \subset Q_L \subseteq Q$. \\
$\mathcal{E}$ & Candidate enrichment types (representation families). \\
$\mathcal{M}_e$ & Generator model options for enrichment type $e\in\mathcal{E}$. \\
$c$ & Active combination (enabled enrichments and their model assignments). \\
$a=(e,m)$ & Atomic unit: an enrichment type paired with a generator model. \\
$\mathcal{U}_\ell$ & Fidelity-local pool of observed atomic units at rung $\ell$. \\
$R(c)$, $\hat{R}(c,\ell)$ & (Mean) Recall@K and its empirical estimate at fidelity $\ell$. \\
$s(c,\ell)$, $\mathrm{SE}(c,\ell)$ & Query-level standard deviation and standard error at fidelity $\ell$. \\
$C_{\mathrm{struct}}(c)$ & Structural deployment cost estimate used for frontier comparison. \\
$C_{\mathrm{uncached}}(c)$ & Runtime uncached dollar spend to evaluate combination $c$. \\
$B_{\mathrm{search}}$ & Search optimization budget limit over uncached spend. \\
$\eta$ & ASHA halving factor controlling promotion quota between rungs. \\
$A_\ell(\kappa)$ & Frontier envelope used for promotion gating and diagnostics. \\
$U_{\mathrm{rank}}(c,\ell)$ & Optional uncertainty-aware ranking score for promotion ordering. \\
$\rho$ & Fraction of chunks evaluated in a working fidelity (proxy for enrichment workload). \\
\bottomrule
\end{tabular}
\end{table}

\section{The CAMI Framework: Atomic Search and Discovery}
\label{sec:method}

To solve the optimization problem formalized above, we introduce CAMI. We design CAMI as a two-phase architecture: a first phase for agentic template discovery, and a second phase for budgeted multi-index search.

\subsection{Multi-Fidelity Evaluation Framework}
Evaluating every candidate combination across the entire corpus is the primary source of search cost. As a solution, our framework restricts offline dollar spend by evaluating combinations at discrete fidelity levels $\ell \in \{0, 1, \dots, L\}$. 

\paragraph{Nested Query Subsets.} Fidelities correspond to nested query subsets $Q_0 \subset Q_1 \subset \dots \subset Q_L$, with fidelity controlling the size of the retrieval-informed working set, thus bounding token costs. For example, with $|Q_0|=20$ and top-$k=10$, retrieval is run for these queries, the retrieved and gold document IDs are unioned, and enrichments are materialized only for chunks in this subset. Let $\hat{R}(c, \ell)$ denote empirical Recall@K of combination $c$ at fidelity $\ell$.

\paragraph{Assumption (Query Log and Relevance Judgments).} CAMI assumes a subset of benchmark query log with gold relevant document IDs for each query. These labels are used strictly for retrieval evaluation and for building the fidelity working sets described above; they are not used to train the retriever. 

\paragraph{Fidelity Representativeness.} Fidelity subsets are sampled from the full query set using topic-aware stratification. This ensures that even small subsets preserve broad coverage and approximate the full query distribution. To monitor whether a subset is a useful screening proxy, the system reports a representativeness score:
\begin{equation}
\mathrm{RepScore}=100\cdot\left(0.4\cdot\mathrm{Coverage}+0.3\cdot(1-\mathrm{JS})+0.3\cdot(1-\mathrm{TV})\right),
\end{equation}
where $\mathrm{Coverage}$ is topic-cluster coverage and the divergence terms are computed between the topic-distribution of the fidelity subset and the full query set. Specifically, $\mathrm{JS}$ denotes Jensen--Shannon divergence and $\mathrm{TV}$ denotes total variation distance:
\begin{equation}
\mathrm{JS}(p,q)=\frac{1}{2}\mathrm{KL}(p\|u)+\frac{1}{2}\mathrm{KL}(q\|u), \qquad u=\frac{1}{2}(p+q),
\end{equation}
\begin{equation}
\mathrm{TV}(p,q)=\frac{1}{2}\sum_i |p_i-q_i|.
\end{equation}
We use $(1-\mathrm{JS})$ and $(1-\mathrm{TV})$ so higher values indicate better fidelity-subset representativeness. 

\begin{algorithm}[t]
\caption{AACA: Atomic-Unit Acquisition + Closure + MO-ASHA}
\label{alg:aaca}
\begin{algorithmic}[1]
\Require Initial/Discovered EDR library $(\mathcal{E},\{\mathcal{M}_e\})$; fidelities $\ell\in\{0,1,\dots,L\}$; budget $B_{\mathrm{search}}$.
\State \textbf{Bootstrap:} Evaluate the content-only baseline at all $\ell$ (excluded from ASHA quotas).
\State \textbf{Bootstrap:} Evaluate deterministic fid0 singleton seeds (cheapest model per $e\in\mathcal{E}$).
\State Initialize trial store, empirical frontiers $\hat{\mathcal{F}}_\ell$, atom pools $\mathcal{U}_\ell$, and pending closure queues.
\State Initialize iteration counter $t\gets 0$.
\While{$\sum C_{\mathrm{uncached}} < B_{\mathrm{search}}$ and time limits permit and $t<T_{\max}$}
    \State Refresh pending closure queues for fidelities with updated atom pools $\mathcal{U}_\ell$.
    \State Query MO-ASHA for the highest-ranked legal promotion (e.g., $0\to1$ or $1\to2$).
    \State Dispatcher selects the highest-ranked action from \{\textsc{promotion}, \textsc{closure}, \textsc{acquisition}\}.
    \If{action is \textsc{promotion}}
        \State Evaluate trial at target fidelity $\ell+1$. Update $\hat{\mathcal{F}}_{\ell+1}$ and $\mathcal{U}_{\ell+1}$.
    \ElsIf{action is \textsc{closure}}
        \State Evaluate highest-priority recombined combination from $\mathcal{U}_\ell$ at fidelity $\ell$.
    \Else \Comment{\textsc{acquisition}}
        \State Admit a new atomic unit $a=(e,m)$ via heuristic/LLM ranking.
        \State Construct legal fid0 combination containing $a$; evaluate at $\ell=0$.
    \EndIf
    \State Update budget ledgers and diagnostics.
    \State Increment iteration counter $t\gets t+1$.
\EndWhile
\State \Return Empirical Pareto frontiers $\hat{\mathcal{P}}_\ell$ and trial history.
\end{algorithmic}
\end{algorithm}
\subsection{Phase 1: Agentic EDR Discovery}
While the search procedure can operate over a statically defined catalog of enrichments (e.g., standard summaries or paraphrases), optimal representations are often highly corpus-specific. We formalize representation generation as a governed proposal function $\mathcal{D}(\mathcal{S},\mathcal{B}) \rightarrow \{\tilde e_1,\ldots,\tilde e_m\}$, where $\mathcal{S}$ is a seed set of query--evidence pairs and $\mathcal{B}$ defines the system's structural constraints.

\paragraph{Governed Template Generation.} The policy model acts as an agentic proposer to expand the representation space. \textit{For example, given a corpus of technical manuals, the agent might propose a \textit{Procedural Extraction} template that instructs the offline generator to convert dense paragraphs into strictly numbered lists.} Each proposed EDR $\tilde e$ includes a semantic brief, a chunk-level generation prompt, and a recommended generator model tier. To maintain a stationary search space, the approved EDRs are deduplicated, schema-validated, and permanently frozen before the inner search begins. Proposer-model token usage is charged to the runtime budget ledger when discovery is enabled, ensuring dynamic representation expansion never bypasses the strict financial accounting of the search phase.

\subsection{Phase 2: The AACA Search Algorithm}

The AACA search phase (Phase 2) evaluates exactly one action per iteration, ensuring stable budget accounting and precise attribution of uncached spend. Algorithm~\ref{alg:aaca} outlines this procedure, which maps directly to the operational stages described below.

\paragraph{Bootstrapping the Search Space.} Rather than immediately searching complex combinations, the system establishes a broad, low-cost empirical baseline (Algorithm~\ref{alg:aaca}, Steps 1--2). The optimizer first evaluates the raw text corpus without enrichments to establish a performance floor, followed by deterministic "singleton" combinations (pairing the base content with exactly one enrichment using its cheapest model). This populates the initial $\ell=0$ atom pool without spending early budget on high-variance proposals.

\paragraph{LLM-Based Event-Driven Atom Acquisition.} The admission of new atomic units explores uncharted regions of the structural configuration space (Algorithm~\ref{alg:aaca}, Steps 13--15). We utilize a language model selector that ranks candidate units based on novelty and estimated synergy. Once an atom is acquired, it is evaluated at $\ell=0$. To prevent excessive token burn, atom acquisition is event-driven; the optimizer considers introducing a new enrichment-model pair only when existing candidate pools are exhausted or progress stagnates.

\paragraph{Fidelity-Local Closure Actions.} A closure action is the systematic recombination of individual atomic units that have already proven successful at a specific fidelity (Algorithm~\ref{alg:aaca}, Steps 11--12). Instead of randomly guessing new full combinations, the closure manager exclusively builds new candidate combinations from the verified atom pool at that fidelity. \textit{For example, if the fidelity-1 pool contains a successful summary generated by Llama-3.3-70B and a successful paraphrase generated by Qwen3-8B, a closure action combines them into a single, novel dual-index combination for evaluation at fidelity-1.} This bypasses the inefficiencies of general-purpose sampling, ensuring the algorithm leverages previously paid evaluation evidence to construct highly synergistic multi-index architectures.

\paragraph{Action Dispatch and Proxy Metrics.} During each iteration, the central dispatcher must select between proposing a promotion, executing a closure, or acquiring a new atom (Algorithm~\ref{alg:aaca}, Step 8). It orders these potential actions using ranked heuristics and proxy metrics. A proxy is an approximation or stand-in metric used when the true, full-dataset metric is not yet known. \textit{For instance, the cost incurred on a small fidelity-0 subset serves as a proxy for the true structural deployment cost on the full dataset.} The default dispatcher ranks actions by prioritizing closure combinations with lower structural-delta proxies first, keeping the search aligned with fast budgeted frontier recovery. 

\paragraph{Strictly Gated MO-ASHA Promotions.} To allocate the optimization budget safely, AACA bridges the closure mechanism with a confidence-aware promotion scheduler based on Multi-Objective Asynchronous Successive Halving (Algorithm~\ref{alg:aaca}, Steps 9--10). The transition between fidelities operates under asymmetric strictness:
\begin{itemize}
    \item \textbf{Screening ($0 \dots L-1$):} Early promotions act as broad filters. The optimizer relies on an Upper Confidence Bound (UCB) ranking score to prioritize candidates. UCB is a mathematical formula that balances the average empirical performance of a combination with its uncertainty; it artificially boosts the rank of combinations that have high potential but are currently evaluated on very small, high-variance query subsets. 
    \item \textbf{Budget-Guarded Confirmation ($L-1 \rightarrow L$):} Because maximum-fidelity evaluation consumes the vast majority of the runtime uncached spend, the final promotion relies on an explicit frontier gate. Only combinations residing directly on the empirical Pareto frontier—or within a strict marginal slack tolerance—are permitted to advance to the final evaluation stage.
\end{itemize}

\paragraph{Drift-Aware Bypass Lanes.} Retrieval configurations frequently exhibit fidelity drift, where a combination performs exceptionally well on a small query subset but its performance degrades on the full benchmark. To safeguard against frontier collapse, the scheduler implements a throttled challenger lane. If the rank correlation between two adjacent fidelities drops below a predefined threshold, the strict gate temporarily relaxes. This admits a small, quota-capped fraction of highly cost-efficient, slightly dominated combinations for higher-fidelity confirmation, ensuring the final Pareto frontier remains robust even when early proxy metrics are noisy.

\paragraph{Mathematical Formulation of Promotion Scores.} For promotion ordering, the scheduler uses a UCB-style score:
\begin{equation}
U_{\mathrm{rank}}(c,\ell)=\min\!\left(1,\ \hat{R}(c,\ell)+k\cdot \mathrm{SE}(c,\ell)\right),
\end{equation}
when the standard error $\mathrm{SE}(c,\ell)$ is available and the query subset $|Q_\ell|$ exceeds a minimum threshold; otherwise, it falls back to the empirical mean $\hat{R}(c,\ell)$. Because frontier estimates fluctuate during search, we use this score strictly as an ordering heuristic for already-eligible candidates rather than as a stationary confidence guarantee. To enforce the strict $\ell \rightarrow \ell+1$ frontier gate, we define the local frontier envelope as:
\begin{equation}
A_\ell(\kappa)=\max\{\hat{R}(c',\ell): c'\in\hat{\mathcal{F}}_\ell,\ C_{\mathrm{struct}}(c')\le \kappa\},
\end{equation}
and establish an optimistic screening bound $U_{\mathrm{prom}}(c,\ell)=\hat{R}(c,\ell)+z_\ell\cdot \mathrm{SE}(c,\ell)$. Under the default strict mode, promotions are admitted only if the candidate sits on the empirical frontier, falls within slack tolerances, or if its optimistic bound remains capable of challenging the envelope under a capped bypass lane. 

\begin{table}[t]
\caption{Processed evaluation datasets.}
\label{tab:dataset-summary}
\centering
\footnotesize
\setlength{\tabcolsep}{4pt}
\renewcommand{\arraystretch}{1.06}
\begin{threeparttable}
\begin{tabularx}{\columnwidth}{@{}l X r r r l@{}}
\toprule
Dataset & Corpus & $N$ & $|Q|$ & Rel/q & Split \\
\midrule
BRIGHT-ES   & Science passages      & 121,249 & 116 & 5.04 & Earth Science \\
BRIGHT-Econ & Economics passages    & 50,220  & 103 & 7.77 & Economics \\
BRIGHT-SO   & Technical QA passages & 107,081 & 117 & 4.09 & Stack Overflow \\
StatCan     & Dialogue-to-table     & 5,907   & 553 & 1.57 & English-test \\
NQTables    & Markdown tables       & 169,898 & 959 & 1.01 & Test \\
\bottomrule
\end{tabularx}
\begin{tablenotes}[flushleft]
\footnotesize
\item $N$ is the number of retrievable units in the processed corpus, $|Q|$ is the number of evaluation queries, Rel/q is the mean number of gold relevant units per query, and Split is the evaluation partition used. BRIGHT-ES, BRIGHT-Econ, and BRIGHT-SO denote the Earth Science, Economics, and Stack Overflow subsets of BRIGHT. All counts are taken from the exact local processed files used in our runs.
\end{tablenotes}
\end{threeparttable}
\end{table}

\subsection{Implementation}
CAMI is implemented in Python on top of our multi-index retrieval framework, which supports hybrid dense+sparse retrieval, multiple named indices over a shared document-ID space, and reciprocal rank fusion (RRF) across active index portfolios. CAMI adds two components to this stack: a governed discovery stage for freezing dataset-specific EDR templates, and an AACA search engine that performs just-in-time multi-fidelity evaluation with cache-aware budget tracking. We defer concrete index schemas, retrieval flow, cache layout, and telemetry details to Appendix~\ref{app:implementation}.

\section{Experimental Setup and Evaluation}
\label{sec:protocol}
We evaluate the framework by measuring how reliably and efficiently it recovers high-quality multi-index Pareto frontiers under strict budget constraints, isolating the impact of the AACA optimizer from the raw utility of the underlying enrichments.

\paragraph{Task and Corpus Specification.}
We evaluate retrieval over a diverse set of labeled corpora: three BRIGHT \cite{su2025brightrealisticchallengingbenchmark} subsets (Earth Science, Economics, and Stack Overflow), \textit{StatCanDialogueRetrieval} \cite{Lu_2023}, and a markdown-converted \textit{NQTables} \cite{herzig2021opendomainquestionanswering} retrieval variant. For each dataset, we define a fixed default catalog of EDR families, including summaries, paraphrases, synthetic questions, and mindmaps, together with a tiered pool of generator models available to each family. Unless explicitly overridden, both curated and discovered EDRs are instantiated from the same generator candidate set,
$\{\texttt{qwen3\_8b}, \texttt{qwen3\_235b}, \texttt{llama3\_70b}\}$,
which correspond to the canonical model identifiers \textit{Qwen/Qwen3-8B}, \textit{Qwen/Qwen3-VL-235B-A22B-Instruct}, and \textit{Meta-Llama/Llama-3.3-70B-Instruct}, respectively. At query time, the retrieval pipeline searches each index in the active portfolio independently and merges the resulting ranked lists using reciprocal rank fusion (RRF). Table~\ref{tab:dataset-summary} summarizes the processed benchmark datasets used in the study, while the appendix tables report the default EDR/model catalog and the discovered EDRs evaluated for each dataset.

\paragraph{Fidelity Construction and Representativeness.} As formulated in Section~\ref{sec:method}, we evaluate portfolios under constrained budgets using nested query subsets $Q_0 \subset Q_1 \subset Q_2$. These subsets are constructed via topic-aware stratification to preserve the broad query distribution of the full benchmark. Candidate portfolios and baselines are compared under identical working-set construction rules, ensuring fidelity strictly bounds the amount of enrichment token spend rather than altering evaluation semantics. To monitor fidelity drift, the system logs our RepScore metric across adjacent fidelities.

\paragraph{Budget-Matched Search Baselines.} To demonstrate the efficacy of exploiting atomic substructure, the primary baseline is a budget-matched Random Search optimizer. This baseline searches the identical categorical portfolio space, utilizes the exact same evaluator and cache state, and operates under an identical multi-fidelity MO-ASHA schedule. Both optimizers are strictly constrained by the same runtime uncached budget ($B_{\mathrm{search}}$).

\paragraph{Controlled Ablations.} To isolate the contributions of CAMI’s components, we conduct controlled ablations under a fixed optimization budget. First, we compare AACA to a budget-matched staged random search. Second, we evaluate the acquisition strategy by contrasting the LLM-guided atom selector with a heuristic baseline to assess whether agentic synergy estimation accelerates frontier recovery. Finally, we test reliance on dynamic representations by comparing Pareto frontiers built from static EDR catalogs versus agent-discovered, corpus-specific templates.

\paragraph{Evaluation Metrics and Criteria.} The primary optimization outcome is the quality of the multi-index combinations discovered under a strict search budget ($B_{\mathrm{search}}$). We measure final retrieval quality using full-corpus \textit{Recall@10}. Because our maximum search fidelity ($\ell=2$) utilizes a representative query subset, we select the optimal configurations from the empirical Pareto frontier and re-evaluate them on the complete dataset to report true full-corpus retrieval gains over the content-only baseline ($\Delta_{\text{AACA}}$). To evaluate search efficiency, we track the trajectory of the maximum recall achieved against the structural budget spent, explicitly comparing CAMI's optimization cost to an exhaustive grid-search oracle. Finally, to ensure algorithmic stability and rigorous baseline comparison, we evaluate performance across multiple random seeds, reporting pairwise win/loss outcomes against the budget-matched random search at specific financial constraints (from \$1 to \$4).

\paragraph{Reproducibility Metadata.} Each run logs the seed, search budget, fidelity schedule, model price table, evaluator settings, and active EDR library. With agentic discovery enabled, we also record the identifier of the frozen EDR specification used in AACA Phase 2. Manifests further capture cache state (cold/warm/copied) and retrieval-cache schema/version metadata, ensuring cache-assisted speedups remain auditable without altering frontier semantics.

\section{Results and Analysis}
\label{sec:results}

We evaluate CAMI across benchmark retrieval corpora to assess search efficiency, cost discipline, and its ability to recover high-quality multi-index configurations. Results are organized around four core analytical questions.

% \begin{table}[t]
% \caption{\textbf{Pairwise Win/Loss Outcomes.} Performance of the AACA optimizer compared against a budget-matched random search baseline. Each cell reports the $W/L$ counts across matched random seeds for a specific dataset and search budget. A win ($W$) indicates CAMI achieved a higher full-benchmark Recall@10 for the deployed combination within that strict budget constraint.}
% \label{tab:winrate_budget}
% \centering
% \small
% \setlength{\tabcolsep}{4pt}
% \renewcommand{\arraystretch}{1.05}
% \begin{tabular}{l c c c c c}
% \toprule
% Dataset & \$1 & \$2 & \$3 & \$4 & Overall \\
% \midrule
% BRIGHT-ES   & 4/1 & 5/0 & 5/0 & 5/0 & 19/1 \\
% BRIGHT-Econ & 3/2 & 4/1 & 5/0 & 5/0 & 17/3 \\
% BRIGHT-SO   & 5/0 & 5/0 & 5/0 & 5/0 & 20/0 \\
% StatCan     & 2/3 & 5/0 & 5/0 & 5/0 & 17/3 \\
% NQTables    & 4/1 & 5/0 & 5/0 & 5/0 & 19/1 \\
% \midrule
% All datasets & 18/7 & 24/1 & 25/0 & 25/0 & 92/8 \\
% \bottomrule
% \end{tabular}
% \end{table}

\begin{table}[t]
\caption{\textbf{Pairwise Win/Loss Outcomes.} Performance of the AACA optimizer compared against a budget-matched random search baseline. Each cell reports the $W/L$ counts across matched random seeds.}
\label{tab:winrate_budget}
\centering
\small
\setlength{\tabcolsep}{4pt}
\renewcommand{\arraystretch}{1.05}
\begin{tabular}{l c c c c c}
\toprule
Dataset & \$1 & \$2 & \$3 & \$4 & Overall \\
\midrule
BRIGHT-ES   & \cellcolor{blue!10}4/1 & \cellcolor{blue!15}5/0 & \cellcolor{blue!15}5/0 & \cellcolor{blue!15}5/0 & \cellcolor{blue!12}19/1 \\
BRIGHT-Econ & \cellcolor{blue!5}3/2  & \cellcolor{blue!10}4/1 & \cellcolor{blue!15}5/0 & \cellcolor{blue!15}5/0 & \cellcolor{blue!8}17/3 \\
BRIGHT-SO   & \cellcolor{blue!15}5/0 & \cellcolor{blue!15}5/0 & \cellcolor{blue!15}5/0 & \cellcolor{blue!15}5/0 & \cellcolor{blue!15}20/0 \\
StatCan     & \cellcolor{red!5}2/3   & \cellcolor{blue!15}5/0 & \cellcolor{blue!15}5/0 & \cellcolor{blue!15}5/0 & \cellcolor{blue!8}17/3 \\
NQTables    & \cellcolor{blue!10}4/1 & \cellcolor{blue!15}5/0 & \cellcolor{blue!15}5/0 & \cellcolor{blue!15}5/0 & \cellcolor{blue!12}19/1 \\
\midrule
All datasets & \cellcolor{blue!6}18/7 & \cellcolor{blue!14}24/1 & \cellcolor{blue!15}25/0 & \cellcolor{blue!15}25/0 & \cellcolor{blue!12}92/8 \\
\bottomrule
\end{tabular}
\end{table}

\begin{table*}[t]
\caption{Full-benchmark Recall@10, optimization cost, and deployment cost. Content-only is the baseline content index. AACA selected is the portfolio chosen by budgeted search and then re-evaluated on the full benchmark query set. Oracle upper bound is the best portfolio from exhaustive grid evaluation over the same active catalog, also re-evaluated on the full benchmark query set. $\Delta_{\text{AACA}}$ denotes the gain over content-only.}
\label{tab:headline-scores}
\centering
\setlength{\tabcolsep}{4pt}
\renewcommand{\arraystretch}{1.05}
\footnotesize
\begin{threeparttable}
\begin{tabular}{lrrrrrrr}
\toprule
Dataset & Content-only & AACA selected & $\Delta_{\text{AACA}}$ & Oracle upper bound & AACA search cost$^\dagger$ (\$) & Selected deploy cost$^\ddagger$ (\$) & Grid search cost$^\dagger$ (\$) \\
\midrule
BRIGHT-ES   & 35.1\% & 38.3\% & +3.2 & 39.2\% & 0.77 & 12.74 & 4.46 \\
BRIGHT-Econ & 16.5\% & 18.6\% & +2.1 & 19.1\% & 0.82 & 15.12 & 2.39 \\
BRIGHT-SO   & 19.8\% & 23.3\% & +3.5 & 23.3\% & 0.73 & 112.13 & 2.62 \\
StatCan     & 30.0\% & 39.4\% & +9.4 & 42.2\% & 3.94 & 9.44 & 13.64 \\
NQTables    & 80.3\% & 87.6\% & +7.3 & 89.8\% & 1.14 & 2.20 & 7.52 \\
\bottomrule
\end{tabular}
\begin{tablenotes}[flushleft]
\footnotesize
\item[$\dagger$] Search costs are measured under the fidelity schedule used during optimization. The highest search fidelity is a representative subset rather than the full corpus, so these values quantify optimization cost under the proxy search protocol rather than the cost of evaluating every candidate on the entire corpus.
\item[$\ddagger$] Selected deploy cost is the full-corpus structural cost of materializing the AACA-selected portfolio once on the final dataset. It is a deployment-time-price and is distinct from cumulative search cost.
\end{tablenotes}
\end{threeparttable}
\end{table*}

\textbf{(Q1) Does the AACA optimizer discover higher-recall portfolios than budget-matched random search?}
Here random search serves as a matched control: it uses the same evaluator, cache conditions, fidelity schedule, and runtime budget, allowing us to isolate the value of atom-aware closure and promotion under identical search constraints. Our findings reveal that the AACA procedure consistently isolates superior combinations compared to this staged random baseline. As detailed in Table~\ref{tab:winrate_budget}, CAMI decisively outperforms the random baseline across all evaluated datasets, achieving an aggregate win rate of 92\% (92 wins to 8 losses) on full-benchmark \textit{Recall@10}. Crucially, the performance advantage widens with the optimization budget. Under tight constraints (\$1), the random baseline occasionally finds competitive configurations, leading to 7 losses for our method. As the budget increases to \$3–\$4, systematic recombination of atomic units lets CAMI explore many more unique multi-index structures at early fidelities without exceeding cost limits. Consequently, AACA achieves a 100\% win rate (25/0) at the highest budget tiers. Because the maximum search fidelity serves as a representative query subset, we validated the selected frontier points on the full dataset, confirming that CAMI’s recovered combinations consistently outperform random sampling given sufficient search capacity.

\begin{figure}[t]
    \centering
    % Replace the filename with your actual image path
    \includegraphics[width=0.9\columnwidth]{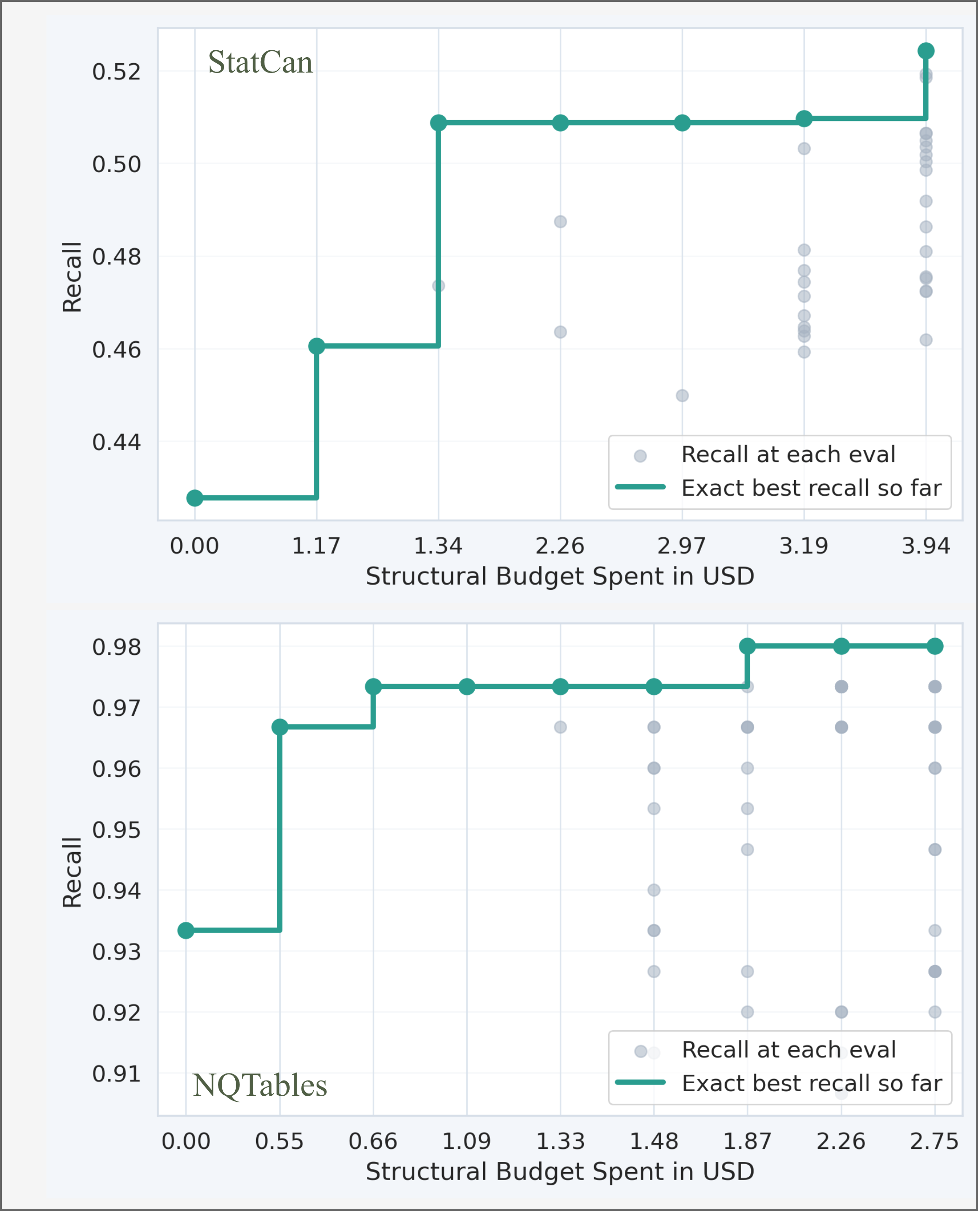} 
    \caption{\textbf{Recall vs. Budget Trajectory at Fidelity-2.} The empirical search progression on the \textit{StatCan} (top) and \textit{NQTables} (bottom) datasets. The solid line tracks the maximum recall achieved (y-axis) as the structural search budget is spent (x-axis). The steep initial ascent demonstrates rapid convergence to near-optimal configurations long before the budget is exhausted. Both get near optimal configurations at less than 20\% (5x less) of the cost of an exhaustive grid search.}
    \label{fig:budget_trajectory}
\end{figure}

\textbf{Search Efficiency and Convergence.}
To quantify cost efficiency, Figure~\ref{fig:budget_trajectory} plots the best recall found versus structural budget at fidelity-2 for \textit{StatCan} and \textit{NQTables}. Both curves rise sharply, reaching near-optimal performance early (under \$2.00). While exhaustive search at this fidelity would cost \$16, our method identifies a highly synergistic portfolio using just 25\% of that budget, achieving recall within 3 points of the optimum. This shows AACA effectively captures the Pareto frontier without the waste of brute-force enumeration. \textit{Notably, even sampled grid search can exceed full deployment cost, motivating efficient search exploration.}

\textbf{Ablation map.}
The Q1--Q4 breakdown is intended to separate the main design decisions in CAMI rather than only compare the full system against one baseline. Concretely, Q1 isolates AACA vs. budget-matched random search, Q2 isolates UCB vs. mean-only promotion, Q3 isolates LLM-guided vs. baseline atom acquisition, and Q4 isolates static vs. discovered EDR catalogs.

\textbf{(Q2) Does uncertainty-aware ranking improve search throughput and frontier quality?}
We evaluated the promotion ranking policy by comparing a strict mean-based approach against the uncertainty-aware Upper Confidence Bound (UCB). Our results show that relying only on the mean score creates a severe bottleneck when early evaluation stages use small, noisy query subsets: genuinely good combinations can receive an unlucky score and fall just short of the strict promotion cutoff. UCB mitigates this by adding the empirical standard error ($\mathrm{SE}(c,\ell)$) to the score, giving high-potential combinations the "benefit of the doubt" so they can survive to the next round of testing. Without this uncertainty buffer, the fidelity $1 \rightarrow 2$ promotion rate on the \textit{StatCan} dataset dropped by 90.6\%. This starved the later evaluation stages of promising portfolios, resulting in a worse final Pareto frontier. Broadly, this is a cost-vs.-confidence tradeoff: increasing the lowest fidelity would reduce noise, but would also make every trial more expensive.

\textbf{(Q3) Does LLM-guided atom selection outperform random acquisition?}
We evaluated the impact of LLM-based selection by comparing it against a random selection baseline. The LLM populated the candidate pool with high-quality atomic units significantly faster than random guessing, avoiding the wasted budget of testing mismatched models and tasks. However, the long-term gains in the Pareto frontier were primarily driven by the fidelity-local closure mechanism rather than by continuous discovery alone. Once a stable pool of useful atoms was established ($\mathcal{U}_\ell$), systematically recombining these proven units yielded much larger improvements than repeatedly asking the LLM for brand new ones. This suggests that discovery expands the candidate library, while closure drives long-run search efficiency once a useful atom pool exists.

\textbf{(Q4) Do dynamically discovered EDR templates yield measurable gains over static catalogs?}
The discovery loop consistently produced corpus-specific EDR templates that were retained for downstream evaluation. The approved discovered sets include \textit{key\_factors} and \textit{temporal\_causal\_relationships} for BRIGHT-Earth Science, \textit{causal\_links} and \textit{entity\_descriptions} for BRIGHT-Economics, \textit{function\_names} for BRIGHT-Stack Overflow, \textit{metadata\_extraction} and \textit{implicit\_facts} for NQTables, and \textit{temporal\_facts} and \textit{statistical\_facts} for StatCan. For details, please refer to table \ref{tab:discovered-edrs}. In AACA Phase 2 runs, most of these discovered units survive to the final $\ell=2$ frontier rather than being immediately dominated by the seed catalog of EDRs. For example, \textit{causal\_links} appears in 2 of the 4 final $\ell=2$ frontier configurations on BRIGHT-Economics, while \textit{function\_names} appears in 3 of the 7 final $\ell=2$ frontier configurations on BRIGHT-Stack Overflow. More concretely, on BRIGHT-Economics, \texttt{content + paraphrase(qwen3\_8b)} achieved accuracy 0.181 with structural search cost 0.044, while \texttt{content + causal\_links(qwen3\_8b)} achieved the same accuracy 0.181 with lower structural search cost 0.038 and therefore landed on the frontier. Because selected views are later materialized over the full corpus, even modest per-view cost differences can matter at deployment scale. This indicates that agentic EDR discovery can expand the representation library with dataset-conditioned views that are competitive on either cost, quality, or both under budgeted search.

\begin{comment}
\section{Limitations}
\label{sec:limitations}

While the CAMI framework effectively isolates high-quality multi-index portfolios, it operates under specific structural constraints. Fundamentally, the AACA optimizer assumes that synergistic portfolios can be incrementally constructed from individually successful atomic units. This approach permanently discards atoms that fail initial low-fidelity screening, potentially missing optimal configurations governed by extreme epistasis. Furthermore, the system's cost-efficiency relies heavily on small query subsets ($Q_0$) serving as accurate proxies for full-corpus performance; extreme fidelity drift can cause the optimizer to waste high-fidelity budget confirming false positives. Consequently, the MO-ASHA frontier gates function as practical empirical heuristics rather than providing formal mathematical guarantees of global Pareto optimality. Finally, candidate acquisition quality depends strictly on the underlying policy model's instruction-following capabilities, and the observed recall--cost trade-offs may require recalibration for radically different data modalities.
\end{comment}

\section{Conclusion, Limitations and Next Steps}
\label{sec:conclusion}

Modern multi-index retrieval reduces semantic mismatch but introduces a combinatorial cost in index construction. We present CAMI, which formulates this as a budgeted multi-objective search problem. By leveraging reusable atomic substructures in retrieval enrichments, the AACA optimizer decouples evaluation from recombination. Empirically, fidelity-local closure with gated MO-ASHA promotions enables efficient allocation of offline computation, recovering high-quality recall–cost Pareto frontiers under tight budgets. Across benchmarks, CAMI achieves up to \textcolor{black}{9.4\%} recall@10 gains while using up to \textcolor{black}{5×} less budget than random search.

While CAMI isolates high-quality multi-index portfolios, it relies on structural assumptions that leave room for improvement. It may miss higher-order interactions when atoms are not admitted, not co-evaluated at the same fidelity, or reached too late within budget. Although closure recombination adds no LLM cost, it still depends on scheduler exploration. Efficiency also depends on proxy subsets ($Q_0$) accurately reflecting full-corpus behavior; fidelity drift can waste budget, making the MO-ASHA frontier heuristic rather than globally optimal. More broadly, performance is bounded by the acquisition policy, variability in LLM outputs, and deployment constraints (e.g., latency, storage). Future work includes modeling stronger non-additive interactions, improving promotion and stopping rules, designing more predictive fidelities, and incorporating additional deployment objectives.
% While CAMI effectively isolates high-quality multi-index portfolios, it relies on structural assumptions that leave room for improvement. The remaining higher-order interaction risk is more constrained than generic combinatorial explosion: CAMI may still miss combinations involving atoms that are never admitted, do not co-occur at the same fidelity, or are reached too late in the closure queue before the budget is exhausted. Although closure recombinations incur no additional LLM-generation cost once atoms are materialized, they still depend on the scheduler exploring them. Cost efficiency also hinges on small proxy query subsets ($Q_0$) accurately predicting full-corpus behavior; fidelity drift can waste high-fidelity budget, making the MO-ASHA frontier an empirical heuristic rather than a guarantee of global Pareto optimality. More broadly, portfolio quality is bounded by the acquisition policy’s instruction-following ability, regenerated LLM outputs can shift the candidate frontier, and deployment constraints (e.g., storage or latency) may require extending the recall–cost objective. Future work includes capturing stronger non-additive interactions, designing more principled promotion and stopping rules, improving fidelity construction via adaptive or more predictive proxies, and incorporating additional deployment objectives when operationally binding.

\bibliographystyle{ACM-Reference-Format}
\bibliography{references}

\appendix

\section{Appendix: Cost Accounting and Token Pricing Model}
\label{app:budget-token-cost}
To calculate the real-world dollar cost for any LLM invocation—whether for generating an EDR text block or executing an agentic policy call—we utilize a standard token-based pricing model. Let $t_{\mathrm{in}}$ and $t_{\mathrm{out}}$ denote the number of input (prompt) and output (completion) tokens for a given request. Given a model-specific price per token $p_{\mathrm{in}}$ and $p_{\mathrm{out}}$, the cost of a single generation request is:
\begin{equation}
C_{\mathrm{req}} = (t_{\mathrm{in}} \cdot p_{\mathrm{in}}) + (t_{\mathrm{out}} \cdot p_{\mathrm{out}})
\end{equation}

For the \textbf{Runtime Uncached Spend} ($C_{\mathrm{uncached}}$), the system strictly accumulates $C_{\mathrm{req}}$ for every cache-miss generation and every LLM-based selector prompt executed during the search phase. 

Conversely, the \textbf{Structural Cost} ($C_{\mathrm{struct}}$) ledger charges each atomic unit $j = (e, m)$ by the maximum nominal construction cost observed across fidelities during the run. Let a portfolio $c$ consist of components $j\in c$. For each component, the system maintains a per-run maximum $\overline{C}_j=\max_\ell \hat{C}_j(\ell)$ over the observed fidelities. The structural cost view of a full multi-index portfolio is:
\begin{equation}
C_{\mathrm{struct}}(c)=\sum_{j\in c} \overline{C}_j
\end{equation}
Structural spend during search is updated incrementally by deltas only when some $\overline{C}_j$ increases. This yields a cache-agnostic estimate of search effort, avoiding double-counting the same component when it is first screened at low fidelity and later confirmed at a higher fidelity.

\section{Appendix: Scheduler Details and Bypass Lanes}
\label{app:scheduler}
This appendix summarizes scheduler details that are important for reproducibility but too verbose for the main narrative.

\paragraph{Promotion skeleton.}
For each rung $\ell$, the scheduler counts completed non-baseline trials $n_\ell$ and enforces a minimum evidence threshold $n_\ell\ge n_{\min,\ell}$ before issuing any promotions. The ASHA quota is then computed as
\begin{equation}
\mathrm{quota}_\ell=\max\!\left(1,\left\lfloor\frac{n_\ell}{\eta}\right\rfloor\right),
\end{equation}
and remaining quota subtracts already-issued promotions $p_\ell$.

\paragraph{Gate modes for $1\rightarrow2$.}
\emph{off} admits all eligible fid1 trials to ranking; \emph{strict} admits fid1 frontier points plus near-frontier points within slack; \emph{soft} adds a capped \emph{cheap-dominated} lane, admitting candidates that are slightly worse in recall than some frontier point but materially cheaper.

\paragraph{Bypass lanes.}
When fidelity drift is high (low overlap or low rank correlation between fidelities), the scheduler can allocate a capped \emph{challenger} lane that admits a small number of dominated fid1 trials for fid2 confirmation, and an optional \emph{history-rescue} lane that reconsiders candidates whose relative standing changed sharply across fidelities. These lanes are activated by drift telemetry and remain quota-capped.

In the current baseline, drift can be measured by rank correlation $\rho$ between portfolio recall estimates at adjacent fidelities over their overlap set; a simple drift score is $D=1-\rho$. Challenger capacity is then throttled to increase with $D$ up to a configured cap, and to collapse to near zero when drift is negligible. Candidates in the challenger lane are additionally filtered by fid0 strength and by slack relative to the content baseline at fid1.

\paragraph{Ranking policy.}
When ranking policy is UCB, the scheduler uses $U_{\mathrm{rank}}(c,\ell)$ only when the empirical standard deviation is available and a minimum sample threshold is met; otherwise it ranks by $\hat{R}(c,\ell)$. This keeps uncertainty-awareness local to ordering among already-eligible candidates.

\paragraph{Multi-objective tie-breakers.}
Within a Pareto rank, candidates are tie-broken deterministically using frontier-utility heuristics such as hypervolume contribution \cite{zitzler2002multiobjective} and best-case optimistic improvements, followed by nominal cost and a stable trial-id order. These tie-breakers affect ordering but do not bypass gating.

\paragraph{Scan order.}
Rungs are scanned in either \emph{low-first} or \emph{high-first} order; low-first prioritizes continued screening at fid0 until there is sufficient fid1 evidence, while high-first prioritizes accelerating fid2 confirmations once fid1 is populated.

\paragraph{Dispatcher defaults and promotion guard.}
The default dispatcher mode is closure-first with a promotion-starvation guard. In the current baseline, closure actions are preferred unless the guard fires; the guard is configured to prevent long closure streaks from starving promotions, with defaults that (i) cap consecutive closure actions before forced promotion at 3, (ii) force promotion when a fid2 promotion is available, and (iii) apply the guard only in fid2-available situations. We report these settings in experiment manifests.

\paragraph{Event-driven atom admission defaults.}
Atom admission is event-driven by default: new atomic units are considered only when promotion and closure do not supply a useful action, with an additional stagnation trigger that allows bounded re-entry to atom acquisition after a fixed number of low-progress iterations. The current baseline uses a stagnation threshold of 8 iterations. We report the admission mode and stagnation threshold for every run.

\section{Appendix: Dispatcher Guards and Retrieval Caching}
\label{app:dispatcher-caching}

\paragraph{Dispatcher Starvation Guards.} Because closure combinations can generate large queues of candidate mixtures, a pure closure-first strategy risks stalling higher-fidelity evaluations indefinitely. We address this by enforcing a promotion starvation guard within the central dispatcher. While the dispatcher defaults to prioritizing fidelity-local closure to exploit already-paid atomic admissions, it forces a promotion execution if the consecutive closure streak exceeds a configured threshold. Consequently, high-fidelity confirmation consistently keeps pace with low-fidelity exploration.

\paragraph{Retrieval Result Caching.} Enrichment-cache reuse alone does not eliminate the repeated retrieval and fusion work required when hundreds of overlapping combinations are evaluated. Therefore, CAMI implements retrieval-result caching directly in the evaluator path. We key both atomic retrieval caches and combination-level fusion caches by dataset version identifiers, index identity, query identity, and retrieval parameters. These caches drastically reduce runtime overhead, rendering closure-heavy search practical at larger budgets. Crucially, they do not alter the target optimization objective or the cache-agnostic structural cost semantics.

\section{Appendix: System Implementation}
\label{app:implementation}

\paragraph{Retrieval Framework.}
CAMI is built on a Python MultiIndex retrieval stack centered on four components: \texttt{HybridRetriever}, \texttt{RAGPipeline}, \texttt{MultiIndexRetriever}, and \texttt{UnitxtEvaluator}. \texttt{HybridRetriever} manages a single retrievable collection, including index construction, loading, and dense/sparse/hybrid search. \texttt{RAGPipeline} manages multiple named indices over a shared document-ID space and exposes both single-index and multi-index retrieval. \texttt{MultiIndexRetriever} performs cross-index fusion and reranking, while \texttt{UnitxtEvaluator} computes retrieval metrics from the returned document IDs.

\paragraph{Index Construction.}
Each index stores chunk-level rows keyed by \texttt{chunk\_id}, together with the source \texttt{doc\_id}, the indexed text, a dense embedding, and a sparse BM25-style representation \cite{robertson2009probabilistic}. The dense field is produced with a SentenceTransformer \cite{reimers2019sentence} encoder and indexed in Milvus Lite using \texttt{AUTOINDEX} with cosine similarity. The sparse field is built from tokenized chunk text using BM25-style weights and indexed using \texttt{SPARSE\_INVERTED\_INDEX} with inner-product scoring and the \texttt{DAAT\_MAXSCORE} inverted-index algorithm. Sparse search uses \texttt{drop\_ratio\_search=0.2}. The framework persists the BM25 model and tokenized corpus to disk so that existing collections can be reloaded without re-tokenization or sparse-model reconstruction.

\paragraph{Chunking and Base Corpus Handling.}
The retrieval framework supports recursive, semantic, and Docling-based hybrid chunking \cite{livathinos2025docling}. In the experiments reported here, however, CAMI operates over the processed benchmark chunk files supplied with each dataset. The base corpus is indexed once as a content view, and every active EDR--model pair defines an auxiliary view over the same underlying document collection. All indices therefore share a common \texttt{doc\_id} namespace, which makes cross-index document fusion well defined.

\paragraph{Within-Index Retrieval.}
For a query, \texttt{HybridRetriever} supports dense-only, sparse-only, and hybrid retrieval. In hybrid mode, the query is encoded both as a dense vector and as a sparse BM25-style vector, and Milvus hybrid search combines the two signals using a weighted ranker. The implementation also supports lazy re-initialization of previously built collections, which allows repeated evaluation runs to reuse stored indices directly.

\paragraph{Cross-Index Fusion.}
Given a portfolio of active indices, CAMI retrieves candidates independently from each index and fuses the resulting ranked lists at the document level. In the reported experiments, this is a physical layout choice rather than a requirement of CAMI: the selected portfolio could also be stored in a tagged merged index, provided view identity is preserved and per-view fusion remains available. We use per-view retrieval followed by reciprocal rank fusion (RRF) because raw similarity scales can differ across heterogeneous EDRs. For example, a fluent summary view may yield higher absolute cosine scores than a keyword-style or mindmap view even when the latter ranks the correct document well within that view. Cross-index fusion therefore uses reciprocal rank fusion (RRF):
\begin{equation}
\mathrm{score}(d)=\sum_i \frac{1}{k_{\mathrm{RRF}}+\mathrm{rank}_i(d)},
\end{equation}
with implementation default $k_{\mathrm{RRF}}=60$. When a document is retrieved from multiple indices, the fusion layer aggregates its RRF contributions and retains the highest-scoring chunk payload for downstream inspection.

\paragraph{EDR Materialization.}
The content index is built directly from benchmark chunks. Auxiliary EDR indices are built from generated text associated with the same source chunks. Most EDRs contribute one textual row per chunk, while \texttt{questions} contributes multiple synthetic questions per chunk; these are indexed as separate rows that still map back to the original \texttt{doc\_id}. Discovered EDRs use the same materialization path as built-in EDRs once their prompt specification has been approved and frozen.

\paragraph{Just-in-Time Fidelity Evaluation.}
Search-time evaluation is implemented as a just-in-time procedure. For a fidelity level $\ell$, the evaluator first selects the query subset $Q_\ell$, retrieves top-$k$ candidates from the content index for each query, unions those document IDs with the gold relevant document IDs, and then includes all chunks belonging to that union. This induced chunk set defines the working corpus for fidelity $\ell$. Enrichments are materialized only for this working set, after which the required indices are built or reused and the candidate portfolio is evaluated through the standard multi-index retrieval path.

\paragraph{Metric Computation.}
The evaluator wraps the retrieval function and computes retrieval quality primarily through a Unitxt-based metric pipeline \cite{bandel2024unitxt}. In practice, the logged search traces expose \texttt{match@k}-style retrieval metrics, together with MRR and MAP when requested. A manual fallback path computes per-query recall directly from retrieved and gold document IDs. The evaluation layer also records sample counts and confidence intervals when available.

\paragraph{Caching and Cost Tracking.}
The implementation maintains separate enrichment, retrieval, and fusion caches. Enrichment caches store generated EDR text for previously materialized chunk--EDR--model triples. Retrieval caches store per-query index lookups, and fusion caches store portfolio-level merged rankings. These caches reduce repeated search-time work but do not alter the structural cost comparison between portfolios. Runtime uncached spend records newly incurred generation and policy-model cost during search, whereas structural max-fidelity cost provides a cache-independent estimate of the cost of constructing a candidate portfolio from scratch at the highest configured fidelity. This structural cost should be read as a materialization proxy rather than a complete systems model of storage footprint, indexing latency, query latency, or memory pressure.

\paragraph{Run Metadata.}
Each run writes a manifest containing the random seed, fidelity schedule, active EDR specification, model aliases and canonical model identifiers, pricing manifest, cache provenance, and full trial history. Discovery-enabled runs additionally record the identifier of the approved frozen EDR specification used by the AACA phase. This metadata is used to make both the search space and the execution path auditable. At the same time, independently regenerated LLM outputs can still change discovered templates or materialized EDR text, so stability across regenerated artifacts should be treated as a separate diagnostic from seed-level reproducibility within a fixed run specification.

\section{Appendix: AACA Closure and Atom-Selection Telemetry}
\label{app:aaca-telemetry}
The AACA implementation records telemetry for the three inner-loop mechanisms: closure generation, atom acquisition, and dispatch arbitration. Closure telemetry includes pending-queue sizes, refresh counts, duplicate-elimination counts, and fidelity-wise closure completions. Atom-selection telemetry includes shortlist sizes, heuristic-versus-LLM ranking usage, invalid acquisition candidates, duplicate rejections, and accepted atom acquisitions. Dispatcher telemetry includes forced promotions due to the starvation guard, deferred closure actions, and maximum closure streak length.

These counters are used in ablations to distinguish different failure modes: weak atom acquisition (few useful atoms admitted), closure saturation (many generated closures but few frontier hits), and scheduler conservatism (frontier-competitive candidates blocked by quotas or gate modes). In addition to human-readable summaries, runs report structured telemetry families corresponding to action counts, closure metrics, atom-selector metrics, promotion-guard metrics, direct-fidelity entry counts by origin, and retrieval-cache metrics (when retrieval caching is enabled).

\section{Appendix: LLM Atom-Ranker Prompt and Output Contract}
\label{app:llm}
The optional LLM policy module is used only for atom acquisition. The prompt provides task context, frontier summaries, a neutral history summary, and a shortlist of legal atomic units, then requests a ranked selection (or top choice) in schema-constrained JSON. To remain robust to real model outputs, the parser accepts common wrapper variants (reasoning + JSON object, reasoning + raw JSON list, or fenced JSON) and extracts the final ranking payload while discarding free-form text.

\section{Appendix: Default Catalog and Discovered EDRs}
\label{app:edr-catalog}
This appendix records the default EDR/model catalog used at initialization and the dataset-specific discovered EDRs admitted for evaluation. We keep these tables explicit because they materially affect both frontier quality and search cost.
The deterministic seed EDR set used for fid0 bootstrap is \texttt{summary}, \texttt{paraphrase}, \texttt{questions}, and \texttt{mindmaps}. Their intended roles are: \texttt{summary} (compress core facts; useful for semantic alignment under wording variation, but can over-compress niche details), \texttt{paraphrase} (rewrite with alternate phrasing/synonyms; useful for lexical mismatch, but can drift semantically), \texttt{questions} (generate likely answerable questions; useful for QA-style query matching, but can overfit to generic questions), and \texttt{mindmaps} (structured concept/relationship view; useful for broad multi-concept queries and entity linkage, but can be sparse for very specific fact lookup). In the current implementation, \texttt{summary}, \texttt{paraphrase}, and \texttt{mindmaps} use document-context prompting, whereas \texttt{questions} emits 10 synthetic queries per chunk that are indexed as separate entries mapped to the same document ID.

\begin{table*}[t]
\caption{Default EDR families and generator-model candidates available to the AACA search before optional discovery. In the current runs, the curated default EDRs and discovered EDRs share the same generator candidate set: aliases \texttt{qwen3\_8b}, \texttt{qwen3\_235b}, and \texttt{llama3\_70b}, mapped to \textit{Qwen/Qwen3-8B}, \textit{Qwen/Qwen3-VL-235B-A22B-Instruct}, and \textit{Meta-Llama/Llama-3.3-70B-Instruct}.}
\label{tab:default-edr-catalog}
\centering
\footnotesize
\setlength{\tabcolsep}{4pt}
\renewcommand{\arraystretch}{1.05}
\resizebox{\textwidth}{!}{%
\begin{tabular}{lllp{0.34\textwidth}l}
\toprule
EDR family & Purpose & Granularity & Candidate models (default tiers) & Notes \\
\midrule
Summary (\texttt{summary}) & Core-fact compression for semantic alignment & Chunk & \texttt{qwen3\_8b}, \texttt{qwen3\_235b}, \texttt{llama3\_70b} & Document-context generation \\
Paraphrase (\texttt{paraphrase}) & Alternate wording for lexical mismatch & Chunk & \texttt{qwen3\_8b}, \texttt{qwen3\_235b}, \texttt{llama3\_70b} & Document-context generation \\
Synthetic Questions (\texttt{questions}) & QA-style query matching anchors & 10 indexed entries per chunk & \texttt{qwen3\_8b}, \texttt{qwen3\_235b}, \texttt{llama3\_70b} & 10 synthetic queries/chunk; shared doc ID \\
Mindmaps (\texttt{mindmaps}) & Concept/relationship structure for broad queries & Chunk & \texttt{qwen3\_8b}, \texttt{qwen3\_235b}, \texttt{llama3\_70b} & Document-context generation \\
\bottomrule
\end{tabular}%
}
\end{table*}

\begin{table*}[t]
\caption{Dataset-specific EDRs generated by the agentic discovery stage and admitted for evaluation. Rows listed here passed governance checks and were frozen into the run-specific EDR library before AACA search. All listed rows use the shared generator candidate set \textit{Qwen3-8B}/\textit{Qwen3-235B}/\textit{Llama-3.3-70B}.}
\label{tab:discovered-edrs}
\centering
\footnotesize
\setlength{\tabcolsep}{4pt}
\renewcommand{\arraystretch}{1.05}
\resizebox{\textwidth}{!}{%
\begin{tabular}{llp{0.40\textwidth}l}
\toprule
Dataset & Discovered EDR name & Intended signal & Short description / prompt intent \\
\midrule
BRIGHT (Earth Science) & \texttt{key\_factors} & Driver variables / explanatory factors & Extract factors influencing a described phenomenon or process as natural-language statements. \\
BRIGHT (Earth Science) & \texttt{temporal\_causal\_relationships} & Temporal and cause--effect structure & Extract temporal ordering and causal relationships as natural-language statements. \\
BRIGHT (Economics) & \texttt{causal\_links} & Cause--effect relations in economic narratives & Highlight causal links describing how events/actions lead to other outcomes. \\
BRIGHT (Economics) & \texttt{entity\_descriptions} & Entity-centric factual context & Produce detailed descriptions of key entities and their contextual roles. \\
BRIGHT (Stack Overflow) & \texttt{function\_names} & Function/command syntax anchors & Extract function names and command syntax cues for technical retrieval queries. \\
StatCanDialogueRetrieval & \texttt{temporal\_facts} & Dates, periods, and frequency references & Extract temporal information (dates, periods, frequency) as compact factual sentences. \\
StatCanDialogueRetrieval & \texttt{statistical\_facts} & Numeric and statistical facts & Extract numeric values and statistical statements as compact factual sentences. \\
NQTables (IBM markdown) & \texttt{metadata\_extraction} & Table/document metadata anchors & Extract metadata such as dates, authors, producers, and labels into compact sentences. \\
NQTables (IBM markdown) & \texttt{implicit\_facts} & Deduced table/context facts & Infer implicit facts from the markdown-rendered content and output compact factual sentences. \\
\bottomrule
\end{tabular}%
}
\end{table*}

\section{Appendix: Structural Max-Fidelity Ledger}
\label{app:budget}
The structural ledger charges each (EDR type, generator model) component by the maximum nominal construction cost observed across fidelities during the run, incrementally by deltas. Let a portfolio $c$ consist of components $j\in c$. For each component, maintain a per-run maximum $\overline{C}_j=\max_\ell \hat{C}_j(\ell)$ over observed fidelities. The structural cost view is
\begin{equation}
C_{\mathrm{struct}}(c)=\sum_{j\in c} \overline{C}_j,
\end{equation}
and structural spend during search is updated only when some $\overline{C}_j$ increases. This yields a cache-agnostic estimate of search effort and avoids double-counting the same component when it is first screened at low fidelity and later confirmed at higher fidelity.

\section{Appendix: Operational Workflow (Abstracted)}
\label{app:workflow}
For reproducible runs in the two-layer CAMI setup, the operational workflow is: (i) optional agentic EDR discovery from a fixed seed set, (ii) freeze the approved EDR specification for the run, (iii) optional cache warmup for newly added EDRs and retrieval caches, and (iv) run AACA search on the frozen specification under a fixed budget. This separation keeps the inner search stationary and makes discovery effects auditable.

\end{document}